\documentclass[sigconf]{acmart}

\AtBeginDocument{%
  \providecommand\BibTeX{{%
    \normalfont B\kern-0.5em{\scshape i\kern-0.25em b}\kern-0.8em\TeX}}}

\copyrightyear{2023}
\acmYear{2023}
\setcopyright{rightsretained}
\acmConference[MMAsia '23 Workshops]{ACM Multimedia Asia Workshops}{December 6--8, 2023}{Tainan, Taiwan}
\acmBooktitle{ACM Multimedia Asia Workshops (MMAsia '23 Workshops), December 6--8, 2023, Tainan, Taiwan}
\acmDOI{10.1145/3611380.3628561}
\acmISBN{979-8-4007-0326-3/23/12}

\makeatletter
\gdef\@copyrightpermission{
	\begin{minipage}{0.3\columnwidth}
		\href{https://creativecommons.org/licenses/by/4.0/}{\includegraphics[width=0.90\textwidth]{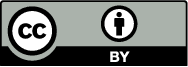}}
	\end{minipage}\hfill
	\begin{minipage}{0.7\columnwidth}
		\href{https://creativecommons.org/licenses/by/4.0/}{This work is licensed under a Creative Commons Attribution International 4.0 License.}
	\end{minipage}
	\vspace{5pt}
}
\makeatother

\usepackage{booktabs}           
\usepackage{multirow}           
\usepackage{amsmath, amsfonts}           
\usepackage{graphicx}           
\usepackage{duckuments}         
\usepackage{subcaption}
\usepackage{xspace}
\usepackage{pifont}
\newcommand{\tool}{\texttt{MMRec}\xspace}
\newcommand{\cmark}{\ding{51}}%
\newcommand{\xmark}{\ding{55}}%

\newcommand{\ie}{\emph{i.e.},\xspace}
\newcommand{\eg}{\emph{e.g.},\xspace}

\newcommand{\etc}{\emph{etc.},\xspace}

\begin{document}

\title[\tool: Simplifying Multimodal Recommendation]{\tool: Simplifying Multimodal Recommendation}

\author{Xin Zhou}
\email{xin.zhou@ntu.edu.sg}
\affiliation{%
  \institution{Alibaba-NTU Singapore Joint Research Institute \\Nanyang Technological University}
  \city{Singapore}
  \country{Singapore}
}

\begin{abstract}
This paper presents an open-source toolbox, \tool for multimodal recommendation. 
\tool simplifies and canonicalizes the process of implementing and comparing multimodal recommendation models. 
The objective of \tool is to provide a unified and configurable arena that can minimize the effort in implementing and testing multimodal recommendation models. 
It enables multimodal models, ranging from traditional matrix factorization to modern graph-based algorithms, capable of fusing information from multiple modalities simultaneously.
Our documentation, examples, and source code are available at \url{https://github.com/enoche/MMRec}.
\end{abstract}




\maketitle

\section{Introduction}

Multimodal recommendation models is a raising trend in current research community due to the following reasons:
\begin{itemize}
	\item The prevalence of multimodal information (\eg images, texts and videos);
	\item Leveraging multimodal information in recommendation can address the sparsity of interaction data~\cite{zhou2023bootstrap, zhou2023enhancing, zhou2022tale, zhou2023comprehensive};
	\item The maturity of recent researches on multimodal learning in NLP and CV domains~\cite{radford2021learning, baltruvsaitis2018multimodal}.
\end{itemize}
Consequently, the multimodal recommendation paradigm has incrementally metamorphosed into an indispensable cornerstone of digital media platforms. This evolution empowers these platforms to deliver tailored recommendations to users. This is achieved by simultaneously scrutinizing historical user interactions and the multifaceted modalities of items~\cite{zhou2023comprehensive}.
Compared with the conventional recommender systems~\cite{zhang2022diffusion, zhou2023layer, zhou2023selfcf, jing2023capturing} that solely leverage user-item interactions for recommendation, multimodal models involves the processes of preprocessing information from multiple modalities (\eg $k$-coring user/item filtering, data splitting, vectorizing multimodal information, aligning items IDs with their multimodal information), fusing multimodal information \etc.
On the one hand, these tedious processes impede the progress of multimodal recommendation research. On the other hand, the wide variety of preprocessing methods make the models difficult to reproduce its performance and compare fairly with others.

To address these issues, we present \tool, an open-source toolbox that simplifies the research on multimodal recommendation. \tool provides a full stack toolbox that includes data preprocessing, multimodal recommendation models, multimodal information fusion, performance evaluation to minimize the cost of implementation and comparison of novel models or baselines. The objective of \tool is to establish a benchmarking system that ensures the fair comparison of multimodal models in an efficient and non-laborious manner.
The toolbox is highly configurable and user-friendly.

\section{Architecture and Key Features}
\begin{figure}[t]
	\centering
	\includegraphics[width=0.48\textwidth]{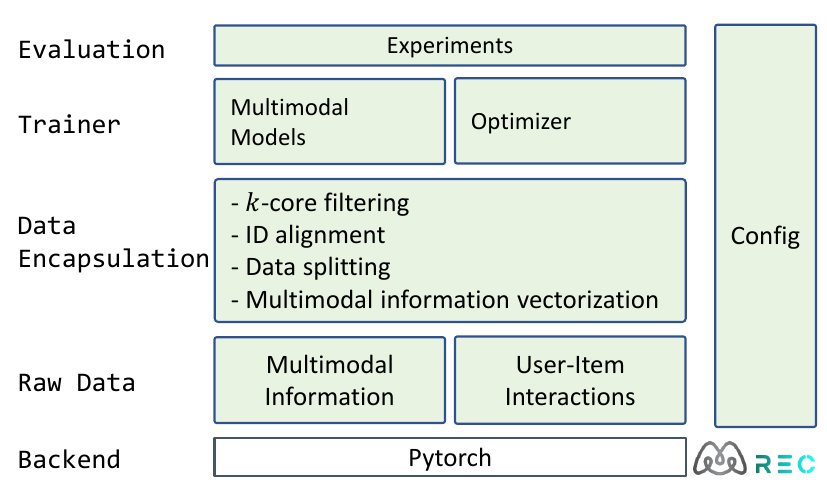} 
	\caption{The architecture of \tool. \tool consists of 4 modules ranging from raw data preprocessing to model performance evaluation. To ensure model reproducibility and performance consistency, \tool consumes raw data as input.}
	\label{fig:framework}
\end{figure}

Figure~\ref{fig:framework} presents the modules of \tool. Noting the inputs of \tool are raw data of multimodal and user-item interaction files.

\textbf{Data Encapsulation.} \tool first preprocesses raw data and encapsulates user interactions and multimodal information into \texttt{DataLoader} of Pytorch. 
The format of raw data is consistent with Amazon Review Data~\footnote{http://jmcauley.ucsd.edu/data/amazon/links.html}. 
\tool performs $k$-core filtering to retain the users and items with at least $k$ interactions, and aligns the multimodal information with the retained items.
It then splits the whole interactions into Training/Validation/Test.
The raw features of multimodal information are vectorized into numeric values leveraging pre-trained multimodal models, such as transformers.

\textbf{Trainer.}
\tool provides various optimizer to train the models. Both unimodal and multimodal models are supported in \tool. In \tool, information from each modality can be easily fused for multimodal recommendation.
In its current version, \tool supports four popular modalities: Text, Image, Audio, Video.
\tool unifies the training interface for all models. 
Customized models are merely required to implement two functions: 

\begin{description}
	\item[\texttt{calculate\_loss}:] The main part of the model, which defines how loss is generated from the model graph flow.
	\item[\texttt{full\_sort\_predict}:] This function predicts the ranking of items for users. 
\end{description}

\textbf{Evaluation.}
This module features a wide set of commonly used metrics for recommender systems, such as Recall, NDCG, MAP \etc.

It is worth noting that all modules can be customized and configured by modifying the configuration files. All changes will be reflected and loaded in \textbf{config} module.
The \textbf{config} module also supports grid searching of models on hyperparameters. The results from all combinations of hyperparameters are summarized and reported to end-users after training.
Reproducibility of models are secured by resetting the seed and raw data in \texttt{DataLoader} in each hyperparameter combination.

\section{Comparison to Related Works}
As far as we know, Cornac~\cite{salah2020cornac} is the only open-source library that supports multimodal information in recommendation.
Although we observe 40+ algorithms are implemented in Cornac, most of them are general recommenders systems that do not utilize multimodal information. 
Furthermore, Cornac is limited by the fusion of modalities in its current stage.
Specifically, Cornac cannot integrate multimodal information (\ie text, image, audio, video) from more than one modality.
Current version of \tool supports 10+ multimodal recommendation models and a board range of general collaborative filtering models.
The detailed comparisons between Cornac and \tool are presented in Table~\ref{tab:comp}.

\begin{table}[bpt]
	\centering
	\caption{Comparison of multimodal recommendation frameworks.}
	\begin{tabular}{l c c c c c}
		\toprule
		\multirow{2}{*}{Frameworks} & \multicolumn{4}{c}{Supported Features} & \multirow{2}{*}{Multimodal Fusion} \\
		\cmidrule{2-5}
		& Text & Image & Audio & Video & \\
		\midrule
		Cornac & \cmark & \cmark & \xmark & \xmark & \xmark \\
		\tool & \cmark & \cmark & \cmark & \cmark & \cmark \\
		\bottomrule
	\end{tabular}
	\label{tab:comp}
\end{table}

\section{Conclusion}
We presented \tool, a multimodal recommendation library to facilitate researchers to implement their models and compare the results with the state-of-the-art baselines. Full documentation and
source code are available at \url{https://github.com/enoche/MMRec}.

\bibliographystyle{ACM-Reference-Format}
\bibliography{sample-base}

\end{document}